\documentclass[%
 reprint,
superscriptaddress,
 amsmath,amssymb,
 aps,
prm,
]{revtex4-1}

\usepackage{graphicx}
\usepackage{epstopdf}
\usepackage{dcolumn}
\usepackage{bm}
\usepackage{gensymb}
\usepackage{upgreek}
\usepackage{hyperref}
\usepackage{tabularx}
\usepackage{chngcntr}
\usepackage{soul}
\usepackage{xcolor}

\preprint{APS/123-QED}

\begin{document}

\title{Human and LLM Collaboration for Accelerated Materials Synthesis and Discovery}

\author{Gregory Bassen}
\email[]{Corresponding author: Gregory Bassen gbassen1@jhu.edu}
\affiliation{Institute for Quantum Matter, William H. Miller III Department of Physics and Astronomy, Johns Hopkins University, 3400 N. Charles Street, Baltimore, MD 21218, United States of America}
\affiliation{Department of Chemistry, Johns Hopkins University, 3400 N. Charles Street, Baltimore, MD 21218, United States of America}
\affiliation{Department of Materials Science and Engineering, Johns Hopkins University, 3400 N. Charles Street, Baltimore, MD 21218, United States of America}
\author{Wyatt Bunstine}
\affiliation{Institute for Quantum Matter, William H. Miller III Department of Physics and Astronomy, Johns Hopkins University, 3400 N. Charles Street, Baltimore, MD 21218, United States of America}
\affiliation{Department of Chemistry, Johns Hopkins University, 3400 N. Charles Street, Baltimore, MD 21218, United States of America}
\affiliation{Department of Materials Science and Engineering, Johns Hopkins University, 3400 N. Charles Street, Baltimore, MD 21218, United States of America}
\author{Sarah Okandey}
\affiliation{Department of Chemistry, Johns Hopkins University, 3400 N. Charles Street, Baltimore, MD 21218, United States of America}
\author{Sarah Cheung}
\affiliation{Department of Chemistry, Johns Hopkins University, 3400 N. Charles Street, Baltimore, MD 21218, United States of America}
\author{Elaine Flowers}
\affiliation{Department of Chemistry, Johns Hopkins University, 3400 N. Charles Street, Baltimore, MD 21218, United States of America}
\author{Ritwik Bose}
\affiliation{Research and Exploratory Development Department, Johns Hopkins Applied Physics Laboratory, 11100 Johns Hopkins Road, Laurel 20723 MD, United States of America.}
\author{Joshua Hummel}
\affiliation{Department of Materials Science and Engineering, Johns Hopkins University, 3400 N. Charles Street, Baltimore, MD 21218, United States of America}
\author{Christopher D. Stiles}
\affiliation{Research and Exploratory Development Department, Johns Hopkins Applied Physics Laboratory, 11100 Johns Hopkins Road, Laurel 20723 MD, United States of America.}
\author{Maxime A. Siegler}
\affiliation{Department of Chemistry, Johns Hopkins University, 3400 N. Charles Street, Baltimore, MD 21218, United States of America}
\author{Tyrel M. McQueen}
\email[]{Corresponding author: Tyrel M. McQueen mcqueen@jhu.edu}
\affiliation{Institute for Quantum Matter, William H. Miller III Department of Physics and Astronomy, Johns Hopkins University, 3400 N. Charles Street, Baltimore, MD 21218, United States of America}
\affiliation{Department of Chemistry, Johns Hopkins University, 3400 N. Charles Street, Baltimore, MD 21218, United States of America}
\affiliation{Department of Materials Science and Engineering, Johns Hopkins University, 3400 N. Charles Street, Baltimore, MD 21218, United States of America}

\date{\today}

\begin{abstract}

Although Large Language Models (LLM) and Artificial Intelligence (AI) tools have enabled a rapid increase in the generation rate of predicted materials, the rate of new materials discovery has lagged behind. This is due to the challenges associated with designing a sequence of chemical reactions to predictably produce new materials, especially in new structure types.  Here, we report a study of human and LLM generated recipes for the synthesis of known and new materials. The success of the recipes is determined through in-lab experimentation, and the results are passed back to the humans and LLMs in a closed-loop process to study the effects of their collaboration. The Ruddlesden-Popper homologous series was selected for all material candidates to provide a materials phase space that is simultaneously well studied and likely to host undiscovered materials. We find that humans (H) and LLM (L) have similar success rates: 83(8)\% (H) and 75(9)\% (L) [known materials, round one], 17(9)\% (H) and 22(10)\% (L) [unknown materials, round one], 79(8)\% (H) and 71(9)\% (L) [known materials, round two], and 22(7)\% (H) and 14(6)\% (L) [unknown materials, round two]. Through this collaborative human-LLM effort, we discovered Ba$_3$PtO$_5$, a material with a new structural prototype that constitutes the missing 1D member of the herein reported dimensionally tunable Rock-Salt Perovskite (RSP) homologous series of the form (AX)$_m$(ABX$_3$)$_p$, of which the Ruddlesden-Popper series is a subset.

\end{abstract}

\maketitle

\section{Introduction}

The current AI/ML revolution has largely been driven by improvements in
computational power, advanced algorithms, and access to large, well-curated datasets \cite{khorshidi_amp_2016,lake_human-level_2015,thompson_lammps_2022,rasmy_med-bert_2021}. This has enabled widespread scientific discoveries, such as in protein structure prediction, exoplanet identification, medical diagnosis, organic retrosynthesis, and many more \cite{jumper_highly_2021, shallue_identifying_2018, mcduff_towards_2025, segler_planning_2018, petrosanu_tracing_2023}. Within materials chemistry and materials physics, AI/ML approaches have greatly accelerated the rate of theoretical material predictions, which now far exceeds the rate of experimental synthesis by orders of magnitude \cite{merchant_scaling_2023}. However, in order for these predicted materials to be useful, they must be synthesizable and experimentally tested. In previous work, we sought to study whether ML tools, particularly when implemented in a closed-loop fashion, could be used for the discovery of superconducting materials \cite{pogue_closed-loop_2023}. We found that uniformly collected experimental data was central to the success of discovering new superconductors in that work.  

Various tools have since been implemented to automate high-throughput synthesis \cite{huang_natural-language-interfaced_2025,szymanski_autonomous_2023,song_multiagent-driven_2025}, but these methods have not considerably changed the overall rate of materials discovery and synthesis, revealing that existing synthesis methods and techniques cannot keep up with the growing corpus of material predictions \cite{elbert_accelerated_2022}. Instead, materials discovery using AI/ML tools is limited by the rate with which materials can be experimentally synthesized and characterized \cite{wilfong_ternary_2025,riesel_crystal_2024}. This grand challenge arises from the fact that to predict synthesizability,
it is necessary not just to represent the ultimate structure of the target material, but to capture all of the important variables controlling the synthetic pathways, i.e. all the factors determining the set of motions atoms undergo during the process of conversion from starting reagents to the final product \cite{bianchini_interplay_2020,mcdermott_graph-based_2021, aykol_thermodynamic_2018}. Moreover, critical factors for inorganic synthesis are incompletely reported in the literature or may not be known. Recent work has used Natural Language Processing (NLP) and Large Language Models (LLMs) to extract structured protocols and to predict synthesizability, precursor sets, and processing conditions \cite{he_precursor_2023,sun_critical_2025,cruse_text_2024,staley_coupling_2026}. Several such tools now exist and quantitative computational benchmarks are emerging; however, prospective experimental validation at scale remains limited \cite{jiang_applications_2025,kim_large_2024,huo_machine-learning_2022,chung_solid-state_2025,prein_language_2025,lee_text-mined_2025,olivetti_data-driven_2020,kononova_text-mined_2019,schrier_pursuit_2023}.

 \begin{figure*}
    \centering
    \includegraphics[height = 8in, width = 7in, keepaspectratio]{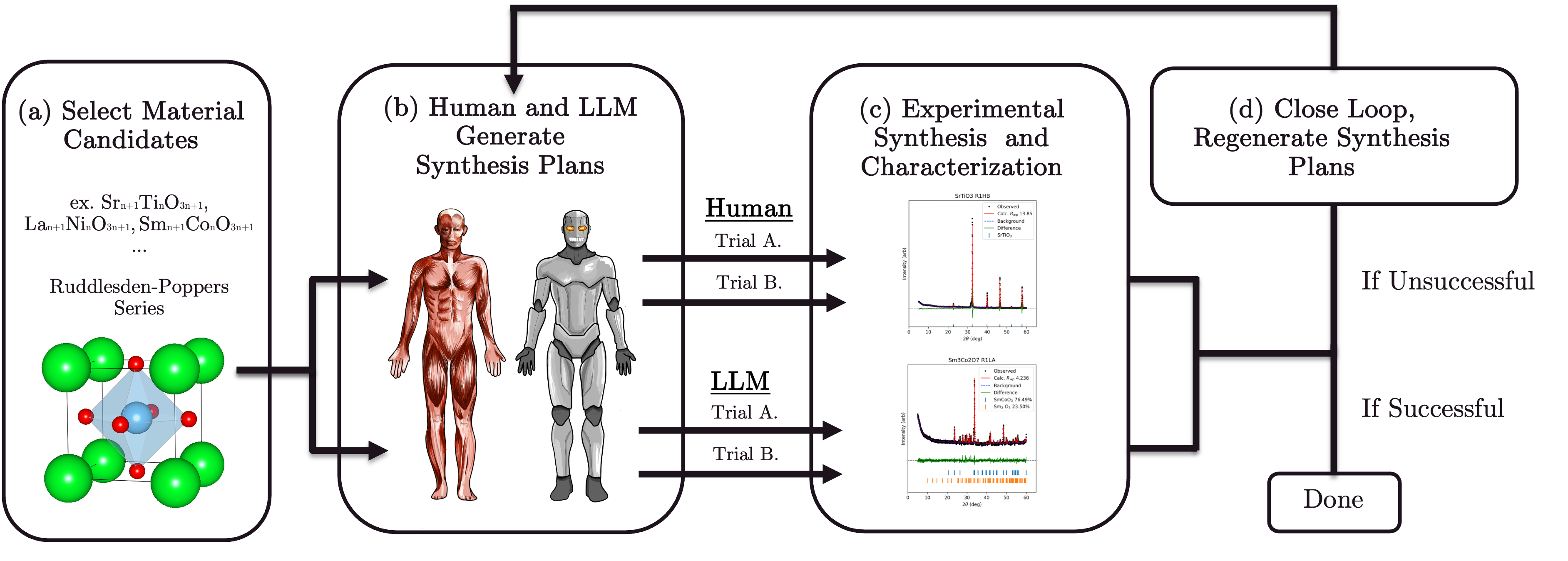}
    \caption{Overall workflow for the synthesis prediction experiment. a) A material candidate from the Ruddlesden-Popper series is selected. This composition may either be known or previously unreported b) both the human and LLM generate synthesis plans c) two trials, A and B, follow each synthesis plan in parallel and characterize the products using powder x-ray diffraction d) if both sets of syntheses were unsuccessful, the results are fed back into the LLM and both the human and LLM regenerate synthesis plans informed from the first round.}
    \label{fig:Workflow}
\end{figure*}

Here we experimentally test LLM and human synthesis recipes both before and after collaboration. The goal is to study the role of collaboration on the synthesis of known materials and the discovery of new ones, while also (i) gathering critical statistical data on LLM recipe performance; (ii) establishing a community standard for model-versus-human assessment; and (iii) seeding creation of a highly curated synthesis database. 

To do so, we select a series of compounds with previously reported members and additional members that are likely to exist but have not yet been realized. We recently demonstrated that new members of a homologous series can be discovered when the ionic radii are proportional to those of known structural analogs \cite{bassen_tolerance_2024}. Here, we apply these structural tolerance factor principles to the Ruddlesden-Popper (RP) homologous series, of the general formula A$_{n+1}$B$_{n}$X$_{3n+1}$ (n = positive integer), where A and B are metal cations and X is an anion \cite{ruddlesden_new_1957,ruddlesden_compound_1958}. These phases contain alternating layers of Perovskite-like ABX$_{3}$ units and Rocksalt-like AX units. Any combination of A, B, and X that are known to form both ABX$_{3}$ (perovskite) and AX, has a high likelihood of stabilizing one or more RP phase. To make syntheses technically straightforward, we set X = oxygen, A = alkaline earth or rare earth, and B = transition metal, with the A and B ions chosen to be in stable oxidation states under ambient synthesis conditions. Targeting previously reported RP members serves to benchmark synthesis success rates for established compounds. Targeting previously unreported RP members, by contrast, enables a direct comparison of human and LLM performance in materials discovery, whether that yields the intended RP phase or an unexpected new phase.

\section{Results and Discussion}

The framework for the synthesis prediction experiment is as follows, and a schematic is shown in Fig. \ref{fig:Workflow}. 
\begin{enumerate}
    \item A target compound is selected.
    \item A human solid-state chemist writes down their synthesis plan, using their knowledge of how materials form. The LLM is asked for a synthesis procedure for the target compound, and its response recorded.
    \item Both synthesis recipes are carried out in two trials, A and B, in parallel, based solely on the written synthesis descriptions.
    \item The synthesis outcomes for each recipe are determined using X-ray diffraction.
    \item If the LLM and human recipes are both unsuccessful, steps 2--5 are repeated. The human  uses the results to design their next reaction, and likewise feed the results back into the LLM to generate another round of synthesis plans. 
\end{enumerate}

This workflow is repeated for different chosen target compounds in order to gather statistically relevant data on LLM performance. A given synthesis plan is carried out in two trials to ensure that written plans contain the relevant information for reproducibility. The final step ensures that this process mimics the actual iterative process of developing a synthesis procedure (series of trial and error), while giving both the model and the human opportunity to learn from each other, providing data on effectiveness of human/AI collaboration. Note that for step 5, if the LLM and/ or human target a previously unreported material and the XRD indicates unidentified peaks, then the purpose of round two is to further isolate the unknown material. 

\begin{figure}
    \centering
    \includegraphics[height = 7in, width = 3.33in, keepaspectratio]{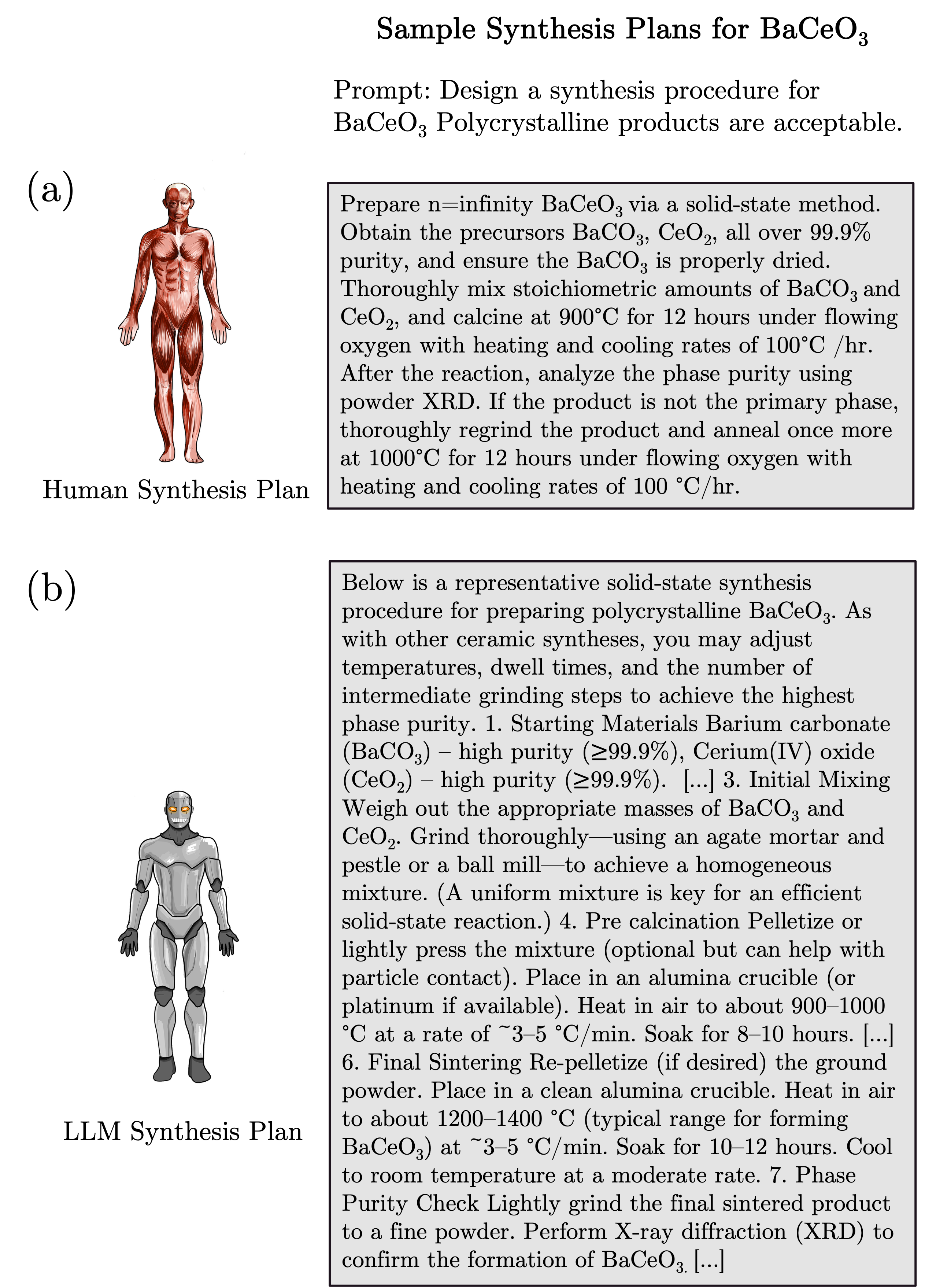}
    \caption{Sample prompt and corresponding synthesis plans. a) Human generated synthesis plan for the targeted material BaCeO$_3$. b) Abridged LLM generated synthesis plan for the targeted material BaCeO$_3$. See SI page S4 and S36-37 for full synthesis plans.}
    \label{fig:Prompt}
\end{figure}

The Human and LLM generated synthesis plans for the targeted material BaCeO$_3$ are shown in Fig. \ref{fig:Prompt} (all other synthesis plans are detailed in the SI). In both BaCeO$_3$ plans, the same starting materials (BaCO$_3$ and CeO$_2$) are selected, thoroughly mixed, heated, and analyzed using XRD. Both follow general standard solid-state synthesis protocol. The LLM plan, however, is noticeably more detailed than the human plan and regularly provides different options for each step. For example, the LLM specifies that the precursors can be ground using either a mortar and pestle or a ball mill, whereas the human plan does not specify how the materials should be mixed. Likewise, the LLM specifies using an Al$_2$O$_3$ crucible, with platinum preferred if it is available, where again the human does not specify the type of crucible to use. The LLM also provides a temperature range for the reaction, rather than discrete temperatures selected by the human. The key synthetic differences between these plans are the atmospheres of the reaction (air for LLM vs O$_2$ for human), the maximum temperature of the second step (1200-1400$\degree$C for LLM vs 1000$\degree$C for human), and the heating and cooling rates (3-5$^\circ$C\ min$^{-1}$ for LLM vs 100$^\circ$C\ hr$^{-1}$ for human). Nevertheless, both plans successfully made phase pure BaCeO$_3$, indicative of the global thermodynamic favorability of BaCeO$_3$ and the large domain of synthesis parameter space that would successfully allow for its formation.

\begin{figure}
    \centering
    \includegraphics[height = 6.6in, width = 5in, keepaspectratio]{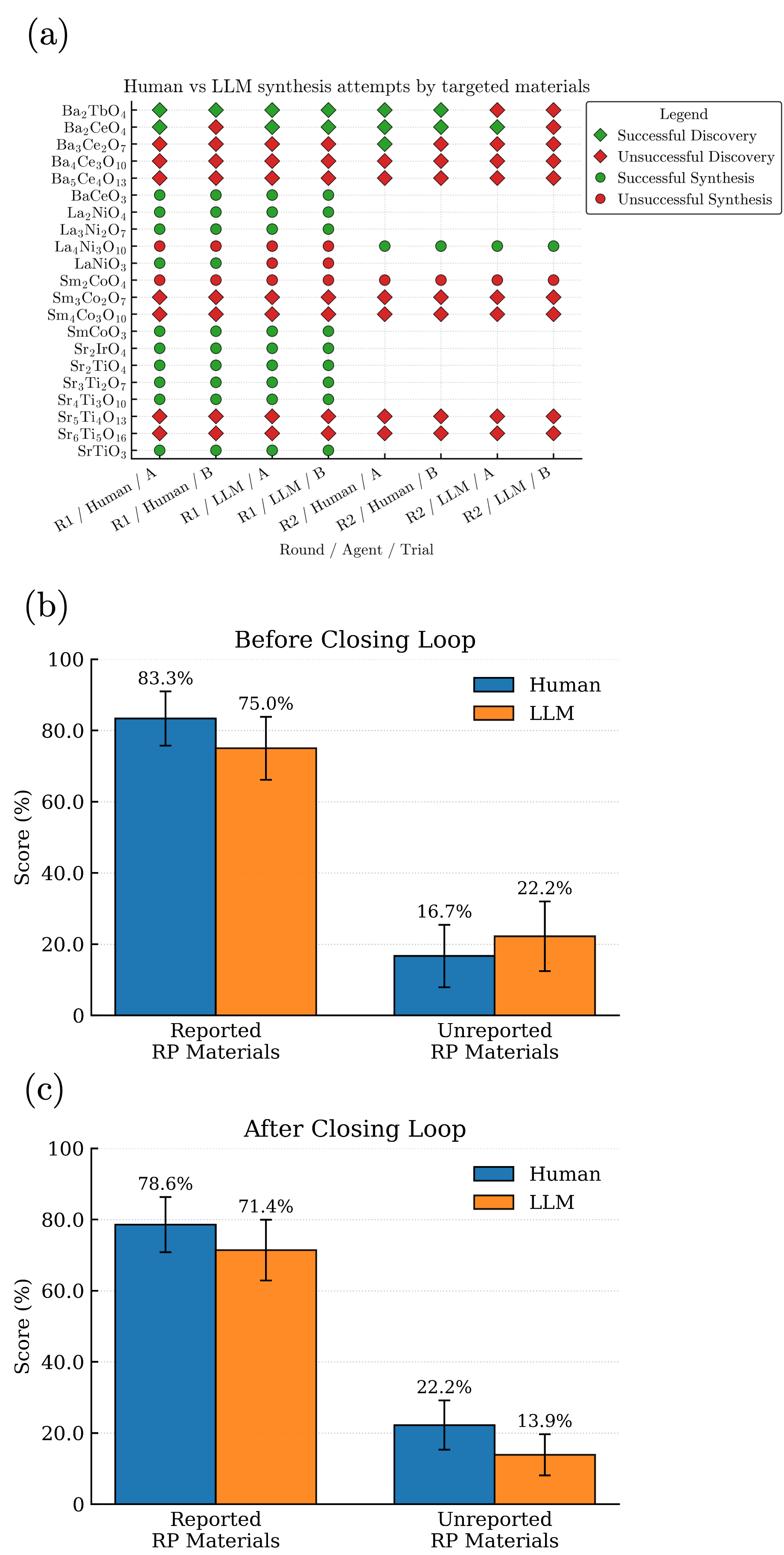}
    \caption{Experimental results of human vs LLM study. (a) Outcome matrix of synthesis attempts. Columns denote round/agent/trial (R1 = before closing the loop; R2 = after closing the loop), agent corresponds to either human or LLM synthesis plan, with A and B indicating independent experimental trials. Rows list target compositions. Circles represent previously reported targets and diamonds represent previously unreported targets as candidates for discovery. Marker color encodes outcome: green indicates a successful result and red indicates an unsuccessful result. For reported targets, success corresponds to formation of the intended phase as determined by PXRD. For unreported targets, success corresponds to observation and discovery of a new phase by PXRD, which may or may not correspond to the originally intended target. (b) Aggregate success rates for reported and unreported RP targets in R1 before closing the loop, comparing human-generated and LLM-generated synthesis plans (c) Aggregate success rates for reported and unreported RP targets in R2 after closing the loop, comparing human-generated and LLM-generated synthesis plans. Error bars denote the standard error of the proportions.}
    \label{fig:Results}
\end{figure}

The experimental results are summarized in Fig.~\ref{fig:Results}a. Our objective with this experiment is to determine which synthesis plans result in the detection of known phases or the discovery of new phases. Accordingly, a reaction targeting a known material is classified as successful if the target is detected by XRD. A reaction targeting a previously unreported material is classified as successful if there is evidence of a discovery. These two test cases are represented by green circles and diamonds respectively. PXRD refinements are shown in SI Fig. S1-S32. 

Before closing the loop, we find that the human and LLM plans allowed for the successful synthesis of several known RP members, which includes BaCeO$_{3}$, La$_{2}$NiO$_{4}$, La$_{3}$Ni$_{2}$O$_{7}$, SmCoO$_{3}$, Sr$_{2}$IrO$_{4}$, Sr$_{2}$TiO$_{4}$, Sr$_{3}$Ti$_{2}$O$_{7}$, Sr$_{4}$Ti$_{3}$O$_{10}$, and SrTiO$_{3}$. The LLM solid state synthesis plan was unsuccessful for the previously reported LaNiO$_3$, whereas the human flux method succeeded. This is due to the high nickel +3 valence state, which requires additional oxygen fugacity. The humans overcame this challenge by using a recently developed flux technique \cite{bassen_mapping_2026}. Moreover, both the human and LLM methods failed to synthesize La$_{4}$Ni$_{3}$O$_{10}$ before closing the loop, as the temperatures and oxygen pressure of their plans were not sufficient to stabilize this structure. However, after closing the loop, both the human and LLM were able to successfully synthesize La$_{4}$Ni$_{3}$O$_{10}$. The human and LLM solid state methods failed to synthesize Sm$_{2}$CoO$_{4}$, both before and after closing the loop. We note that the only reported synthesis of Sm$_{2}$CoO$_{4}$ involved a laser diode floating zone growth under argon gas \cite{lehmann_beitrag_1980}, and our solid state conditions likely did not allow for sufficient reactant diffusion. 

Both the human and the LLM were unsuccessful in synthesizing the higher RP homology candidate materials that have not previously been reported, specifically Ba$_{n+1}$Ce$_{n}$O$_{3n+1}$ (n = 2, 3 and 4) Sm$_{n+1}$Co$_{n}$O$_{3n+1}$ (n=1, 2 and 3) and Sr$_{n+1}$Ti$_{n}$O$_{3n+1}$ (n=4, and 5). We note that although these bulk materials have not been reported, we suspect that the likelihood of making these different candidates are not equivalent. For example, we suspect that the probability of forming the n=4 Ba$_{5}$Ce$_{4}$O$_{13}$ is much lower than of forming n=3 Ba$_{4}$Ce$_{3}$O$_{10}$, as higher nominal RP homologies (n$>$3) are significantly less stable in the bulk and instead tend to form the more thermodynamically favorable perovskite plus lower n RP phases.

There are also notable discrepancies between trials A and B. For example, the A and B trials of R2 LLM Ba$_2$CeO$_4$ resulted in one success and one failure. This can be explained as the LLM provided multiple possible pathways for the synthesis, with trial A following the solid state reaction and trial B following the BaCl$_2$ flux reaction, the former of which was successful and the latter unsuccessful. However, is not clear why R1 human trial A and B results for Ba$_2$CeO$_4$ also differed, as both followed the same synthesis plan. This may reveal the influence of minor differences in reaction conditions in resulting phase formation, such as temperature fluctuations between furnaces, relative humidity between trials, or the quality of the starting materials. Moreover, not all successful syntheses of known materials are equivalent in practice, as reactions produce varying quantities of the target. For example, the synthesis of La$_{3}$Ni$_{2}$O$_{7}$ was successful for both the human and LLM methods, insofar as the targeted phase was detected, but was not the primary phase, suggesting that these synthesis plans need further improvement for phase pure optimization.

Overall, for round one, we find that the human and LLM plans allowed for the successful synthesis of 83(8)\% and 75(9)\% of known RP members, and evidence of discovery in 17(9)\% and 22(10)\%  of unknown RP targets, respectively, as shown in Fig \ref{fig:Results}b. This data suggests that before closing the loop, the LLM does about as well as humans in generating successful synthesis plans for both targeting known materials and discovering new phases (within one standard error of the proportions). After closing the loop, as shown in Fig \ref{fig:Results}c, the human and LLM plans had aggregate success rates 79(8)\% and 71(9)\%  for known RP members, and evidence of discovery in 22(7)\%  and 14(6)\% of unknown RP targets, respectively. The aggregate success rates decreased relative to round one, as there were a higher proportion of unsuccessful syntheses. The human and LLM again performed within one standard error of the proportions when synthesizing previously reported materials, but the human outperformed the LLM in discovering new phases. 

Because Ce and Tb are known to adopt the $+4$ oxidation state in the perovskites BaCeO$_3$ and BaTbO$_3$, we sought the $n=1$ Ruddlesden--Popper (RP) analogues, Ba$_2$CeO$_4$ and Ba$_2$TbO$_4$. For both the human and LLM, the first round products targeting  Ba$_2$TbO$_4$ exhibited unidentified peaks in the PXRD, corresponding to a unit cell with P4/mmm symmetry and  parameters of a=4.29 Å and b=8.76 Å from LeBail refinement, as shown in Fig. S34. Likewise, targeting Ba$_2$CeO$_4$ also resulted in unidentified peaks for several trials, corresponding to a unit cell with P4/mmm symmetry and  parameters of a=4.38 Å and b=13.33 Å from LeBail refinement, as shown in Fig. S35. We do not discuss these structures any further, instead leaving them as a challenge for the community to predict.  

\begin{figure}
    \centering
    \includegraphics[height = 7in, width = 3.33in, keepaspectratio]{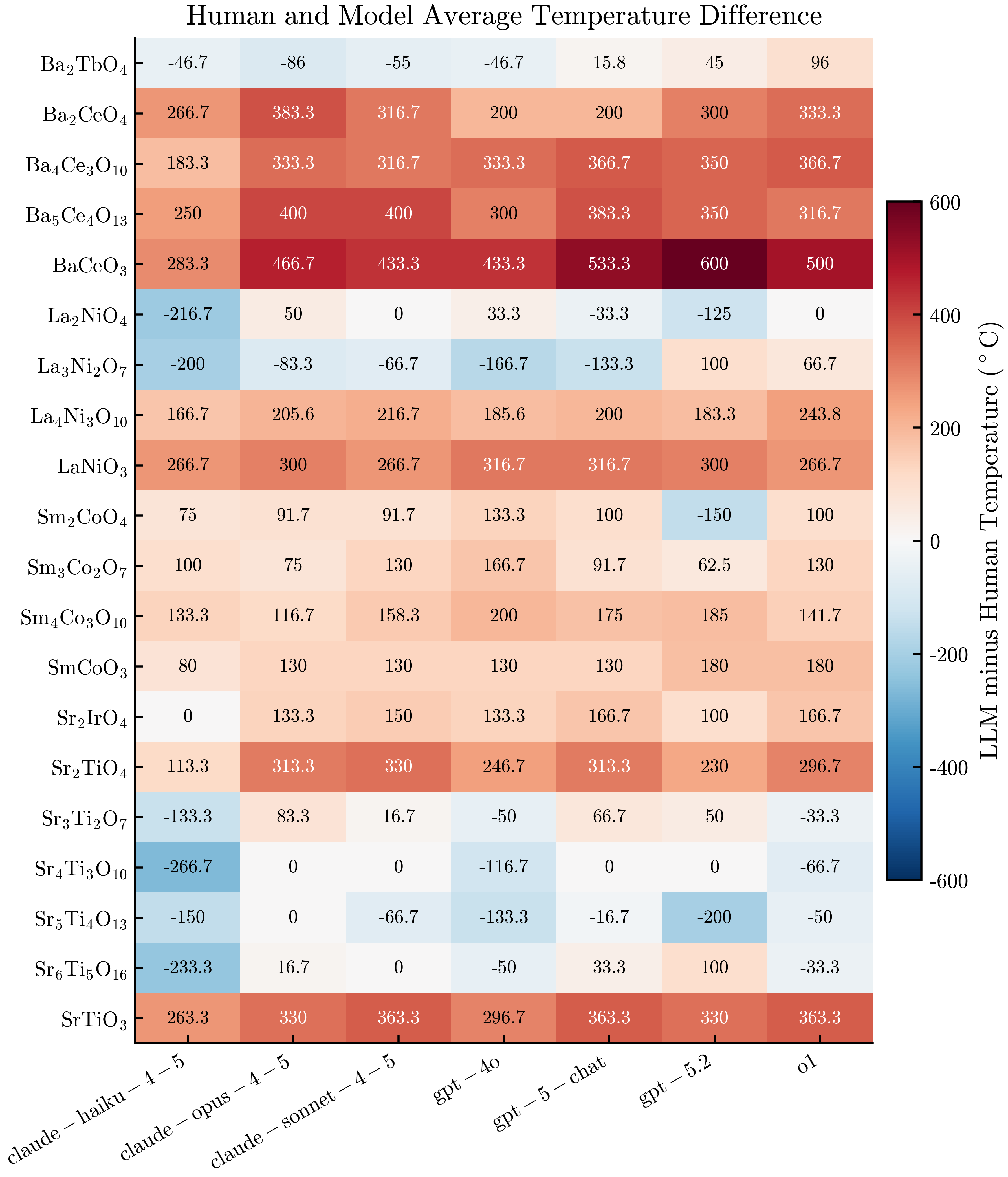}
    \caption{Comparison of human synthesis plans to different LLMs Heat map of the average difference between LLM and human maximum temperatures for a given target material and corresponding synthesis plans.}
    \label{fig:Analysis}
\end{figure}

A comparison of the human synthesis plans with different LLMs is shown in Fig. \ref{fig:Analysis}. As temperature is one of the primary driving thermodynamic factors determining reaction outcome, we plot the difference in the maximum temperature between the LLM and human plans. For materials with two rounds, the average temperature difference is used. There are several models whose maximum temperatures are the same as that of the human plan, such as GPT-5.2 for Sr$_4$Ti$_3$O$_{10}$. We also note that each tested model was within 100$\degree$C of the human plan for Ba$_2$TbO$_4$, which is a previously unreported material and resulted in the discovery of a new phase when experimentally synthesized between 1125$\degree$C-1360$\degree$C. This suggests that each of these models would likely have been successful in discovering this phase. There are also notable discrepancies, such as GPT-5.2 for BaCeO$_3$ which was at a significantly higher temperature than the human method. However, due to the aforementioned favorability of the formation of BaCeO$_3$, it is likely that there is a large temperature range where it could be successfully grown. 

\begin{figure*}
    \centering
    \includegraphics[height = 6in, width = 6in, keepaspectratio]{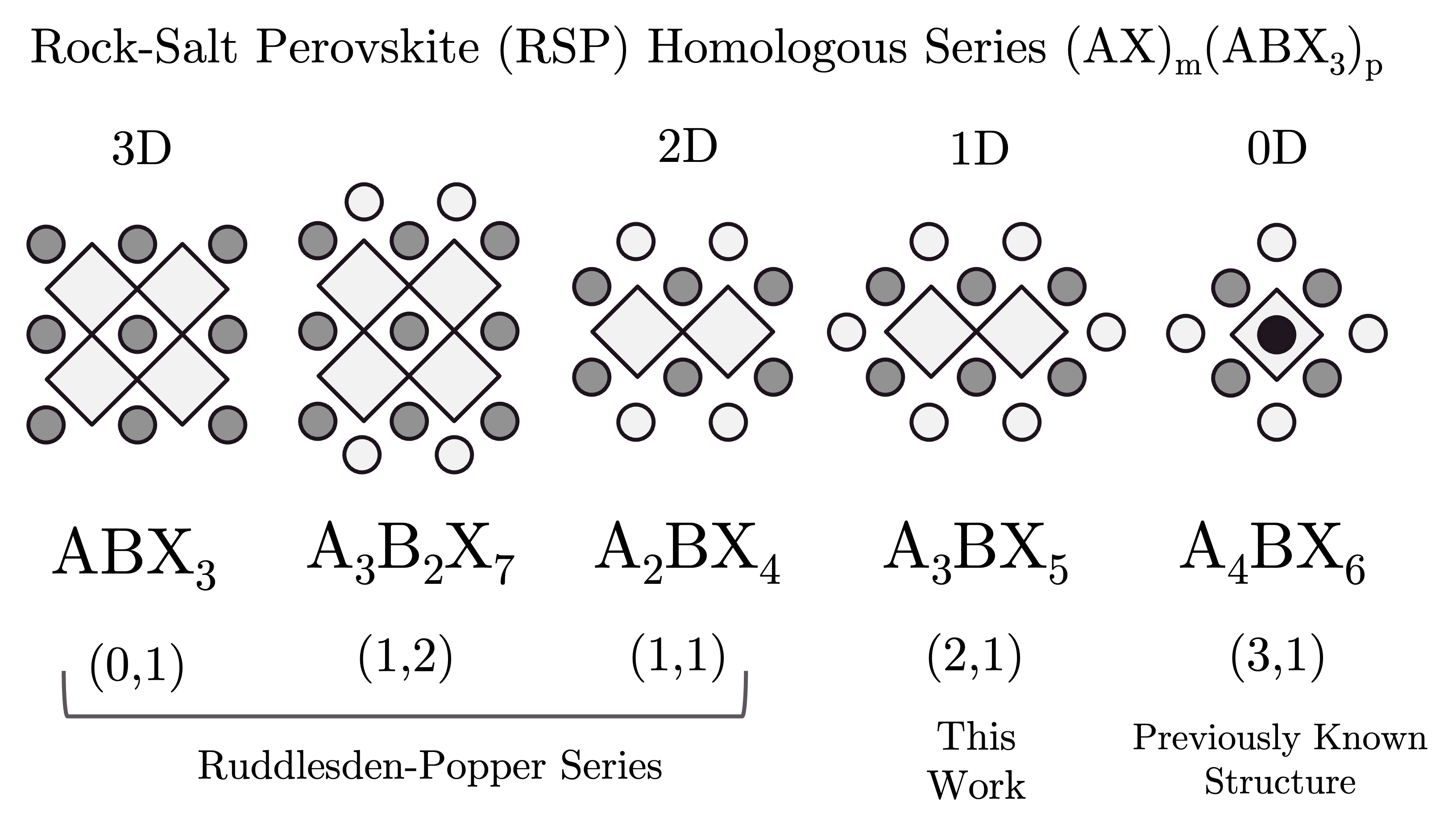}
    \caption{Rock-Salt Perovskite Homologous Series (AX)$_m$(ABX$_3$)$_p$ and the new structural prototype A$_3$BX$_5$. Ruddlesden-Poppers are a dimensionally tunable perovskite homologous series of the form A$_{n+1}$B$_n$X$_{3n+1}$. Members of the Ruddlesden–Popper series correspond to additional perovskite layers inserted between rock-salt layers for each integer value $n$, spanning 3D to 2D connectivity of perovskite units. The classic perovskite $n=\infty$ ABX$_3$, is a continuous 3D network of corner-sharing BX$_6$ octahedra. For $n=1$, A$_2$BX$_4$ consists of 2D perovskite slabs with an additional rock-salt AX layer along one spatial direction. Beyond the Ruddlesden-Popper series, there is also a known ``0D" structure with isolated BX$_6$ octahedra. In this work, we discover the new prototype A$_3$BX$_5$, with one-dimensional chains of corner-sharing octahedra and off-stacked rock-salt units interrupting connectivity along two spatial directions. This finding connects the 2D RP to the 0D case, forming a super family which we call the Rock-Salt Perovskite Homologous Series (AX)$_m$(ABX$_3$)$_p$, relating the number of rock-salt and perovskite units r and p respectively. The classic 3D perovskite can now be described as $m=0$, $p=1$; 2D RPs are $m=1$, $p\geq1$; the new 1D prototype A$_3$BX$_5$ is $m=2$, $p=1$; the 0D case is $m=3$, $p=1$.}
    \label{fig:Series}
\end{figure*}

Whereas most synthesis prompts did not specify single crystal growth, the observation of unknown peaks in Ba$_2$TbO$_4$ prompted us to target single crystals in round two. The human plan (SI Human Round Two Ba$_2$TbO$_4$ page S43) involved a flux growth using BaCO$_3$ and K$_2$CO$_3$ with Tb$_4$O$_7$ in a Pt crucible. Instead of the intended reaction between BaCO$_3$ and Tb$_4$O$_7$, BaCO$_3$ reacted with the Pt crucible, yielding single crystals and the serendipitous discovery of Ba$_3$PtO$_5$. The Ba$_3$PtO$_5$ structure was solved by single-crystal X-ray diffraction (SCXRD); crystallographic details are provided in the SI Tables S1 and S2. To the best of our knowledge, Ba$_3$PtO$_5$ represents the first corner-sharing octahedra one-dimensional motif, built from perovskite and rock-salt units, as seen in SI Fig S33. Further, Ba$_3$PtO$_5$ is related to the 1D lead halide perovskite structures with functionalized molecules in the A site \cite{daub_1d_2021,daub_synthesis_2025}. The connectivity reported herein is suggestive of the dimensional reduction achieved in the $n=1$ RP structure A$_2$BX$_4$, where corner-sharing octahedral sheets are separated by rock-salt layers. More generally, we propose a continuous dimensional-reduction sequence in a perovskite-derived homologous series (AX)$_m$(ABX$_3$)$_{p}$: three-dimensional connectivity in ABX$_3$ with $m=0$ and $p=1$, two-dimensional connectivity in the $n=1$ RP phase A$_2$BX$_4$ with $m=1$ and $p=1$, one-dimensional connectivity in the newly identified A$_3$BX$_5$ motif with $m=2$ and $p=1$, and zero-dimensional connectivity in the 0D reported A$_4$BX$_6$ structure type where $m=3$ and $p=1$. Fig.~\ref{fig:Series} illustrates this progression, in which successive insertion of rock-salt layers interrupts the corner-sharing octahedral network along one additional crystallographic direction. The inclusion of p, the number of perovskite units relative to rock salt, accounts for the Ruddlsden-Popper members between ABX$_3$ and A$_2$BX$_4$, such as A$_3$B$_2$X$_7$ where $m=1$ and $p=2$. All RP members exist as $m=1$ and increasing integer values of p up to the perovskite where $m=0$ and $p=1$.

In addition to Ba$_3$PtO$_5$, the targeted Ba$_2$TbO$_4$ exploratory flux grown human sample also led to the serendipitous discovery of H$_2$BaPt$_2$Cl$_{12}$$ \cdot 9 $(H$_2$O). Here, crystals were discovered after routinely cleaning the Pt crucible in HCl and inadvertently leaving the system undisturbed for the acid to evaporate. The crystals were then analyzed using SCXRD and the structure is shown in SI Fig. S36. This otherwise routine crucible cleaning step was not included in the the human recipe, and thus serves as an example of implicit domain expertise knowledge leading to discovery.

\section{Methods}
All samples were prepared according to their respective synthesis plan, which are shown in the SI (pp. 1-80). Ba$_{n+1}$Ce$_{n}$O$_{3n+1}$ (n = 1,2,3,4 and $\infty$) were targeted with BaCO$_3$ (Alfa Aesar, Q18C024) and CeO$_2$ (Alfa Aesar, F22Y027). La$_{n+1}$Ni$_{n}$O$_{3n+1}$ (n=1,2,3 and $\infty)$ were targeted with La$_2$O$_3$ (Beantown Chemicals, 50001777) and NiO (STREM Chemicals, Inc., L02402203). Sm$_{n+1}$Co$_{n}$O$_{3n+1}$ (n=1,2,3 and $\infty)$ were targeted with Sm$_2$O$_3$ (NoahTech, 48060) and Co$_3$O$_4$ (Beantown Chemical, 50004607). Sr$_2$IrO$_4$ was targeted with SrCO$_3$ (Beantown Chemicals, 50000729) and IrO$_2$ (J\&J Materials, D24666). Sr$_{n+1}$Ti$_{n}$O$_{3n+1}$ (n = 1,2,3,4,5 and $\infty$) were targeted with SrCO$_3$ (Beantown Chemicals, 50000729) and TiO$_2$ (Alfa Aesar, T05D048). Ba$_2$TbO$_4$ was targeted with  BaCO$_3$ (Alfa Aesar, Q18C024) and Tb$_4$O$_7$ (Alfa Aesar, W09F004).

Powder X-ray diffraction patterns were collected at room temperature on ground powders of each sample using a Bruker D8 Focus or  Bruker D8 Advance  powder diffractometer with Cu K$_\alpha$ radiation, 0.6mm divergence slit, 6mm detector slit, 2.5$\degree$ detector Soller slit, Ni 0.5mm and Ni 0.02mm K$_\beta$ filters, and LynxEye and LynxEye XE-T detectors respectively. The X-ray diffraction patterns were measured in the range 2$\theta$=5$^{\circ}$-60$^{\circ}$ with a step size of 0.01715$^{\circ}$ and a collection time of 15 minutes each. Automated Rietveld refinements were performed using GSAS \cite{larson_general_1994}. 

Single crystal X-ray diffraction data for Ba$_{3}$PtO$_{5}$ were collected at T = 213.00(10) K on a Rigaku XtaLAB Synergy-R diffractometer equipped with a rotating-anode X-ray source and HyPix-6000HE detector using Cu K$\alpha$ radiation ($\lambda$ = 1.54178 \AA). Cell refinement and data reduction were performed in CrysAlisPro (Version 1.171.42.49, Rigaku OD, 2022). The structure was solved with SHELXT-2018/2 and refined on $F^{2}$ with SHELXL-2019/3 \cite{sheldrick_crystal_2015}. A semi analytical absorption correction using a multifaceted crystal model was applied in CrysAlisPro. Temperature control was maintained with an Oxford Cryosystems Cryostream 1000.

\section{Conclusion}
In conclusion, we find that human and LLMs perform similarly in the synthesis of known materials and discovery of new materials when targeting Ruddlesden-Popper oxides. Moreover, the collaboration between human and LLM led to the discovery of Ba$_3$PtO$_5$, a novel 1D perovskite derived structural prototype. These results indicate that collaboration with LLMs can be useful for the synthesis planning of known materials and for the guided exploratory synthesis and discovery of new materials.

\section{Acknowledgments}

 We would like to acknowledge Carly Weisblum for illustrating the anatomical `universal human' and LLM robot in Figures 1 and 2.

\section{Funding}

CDS and RB gratefully acknowledge internal financial support from the Johns Hopkins Applied Physics Laboratory's Independent Research \& Development (IR\&D) Program for funding portions of this work. SC, EF, and SO acknowledge support from the David and Lucile Packard foundation through the FiGURE program. GB acknowledges support from the Harry and Cleio Greer Fellowship. Financial support from the NIH (Grant No. 1S10OD030352) awarded to M.A.S. is gratefully acknowledged. 

\section{Author Contributions}

G.B. and T.M.M. contributed to the conception of the work. G.B. designed all human synthesis plans and generated the LLM plans. S.O., S.C., E.F., G.B., W.B., and J.H. synthesized materials and collected experimental PXRD data. M.A.S. collected the SCXRD data. M.A.S., G.B., and T.M.M., analyzed the SCXRD data and solved the structure for Ba$_3$PtO$_5$. W.B. and G.B. analyzed the PXRD data and plotted the refinements. G.B. and T.M.M. contributed to the writing and revising of the manuscript. R.B., G.B., and C.D.S. compared human synthesis plans to different LLMs. 

\section{Competing interests}
There are no competing interests to declare.

\section{Data and Materials Availability}

All samples synthesized in this experiment are stored in the McQueen Lab at Johns Hopkins university. All data will be made available in an online data repository. 

\appendix*

\counterwithin{figure}{section}

\bibliography{references}
\bibliographystyle{achemso}

\end{document}